# Induced energy polarization of the vacuum and the resulting cosmology

Author:          A. Raymond Penner

Address:         Department of Physics,

                 Vancouver Island University,

                 900 Fifth Street,

                 Nanaimo, BC, Canada,

                 V9R 5S5

Email:           raymond.penner@viu.ca

Tel:             250 753-3245 ext: 2336

Fax:             250 740-6482

**Induced energy polarization of the vacuum and the resulting cosmology**

**Abstract**

The theory of an induced energy polarized vacuum provides an alternative to the standard cosmological model. The theory has previously been shown to lead to the Baryonic Tully-Fisher Relationship [1], to agree with the observed rotation curve of the Galaxy [2], to readily produce the observed features in the rotational curves of other spiral galaxies [3], and to agree with observations of the Coma cluster [4]. All without the need for any free parameters. The theory of an induced energy polarized vacuum is now applied to superclusters. From a model of the distribution of superclusters, the overall density parameter of universe as given by the theory of an induced energy polarized vacuum is $\Omega = 0.94 \pm 0.23$. This is consistent with a geometrically flat universe. In addition, by treating the superclusters as unbound systems, the theory leads to an accelerating expansion of the universe in line with observations and without requiring the need for dark energy.

## 1. Introduction

In the standard $\Lambda$CDM model of cosmology, the three primary energy components of the universe are ordinary baryonic matter, dark matter, and dark energy which is equivalent to a cosmological constant. Dark matter is usually taken to consist of currently undetected non-baryonic particles that interact only via the weak force and gravity. The prime candidate for the possible dark matter particles are weakly interacting massive particles (WIMPs) predicted by supersymmetry models. Dark energy is, most commonly, taken to be a constant vacuum energy density that fills all space. This standard cosmological model and the relative contributions that each of the three constituents makes to the overall energy content of the universe stems primarily from the following observations and theoretical considerations.

i)     From models of Big Bang nucleosynthesis and observations of the abundance of the light elements, the current density parameter for baryons, $\Omega_{B,O}$, is estimated to be [5]

$$\Omega_{B,O} = (0.020 \pm 0.002)h^{-2} \tag{1}$$

where Hubble's constant is expressed as $H_O = (100 \text{ km s}^{-1} \text{ Mpc}^{-1})h$. For a value of

$$H_O = (73.8 \pm 2.4) \text{ km s}^{-1} \text{ Mpc}^{-1} \tag{2}$$

as given by Riess et al [6] the current density parameter for baryons will therefore by (1) be

$$\Omega_{B,O} = 0.037 \pm 0.006 \tag{3}$$

It is estimated that approximately 10% of the baryonic mass resides in stars while the rest is primarily found in large gas clouds such as those that dominate the baryonic mass of large clusters.

ii)    The rotational velocities of stars and gas clouds in the outer regions of spiral galaxies are found to be much greater that what is predicted from the baryonic mass of these galaxies. Examples of galactic rotational curves are provided by Sofue [7-8] and Nordermeer [9]. In general, the rotational velocities are found to approach a constant value in the outer regions with little indication that they will eventually fall off in a Keplerian fashion. It is difficult to determine exactly how much additional mass is needed. However, it would appear that an order of magnitude more mass than what is provided by the baryonic mass is required.

       Also the velocity dispersions of galaxies in galactic clusters indicate that approximately an order of magnitude more mass than what is provided by the baryonic mass is required. Indeed, this result is what first led to the idea of dark matter [10]. The additional mass required by clusters is confirmed by shear measurements and by observations of the X-ray emission of the intra-cluster gas which is the dominant baryonic component of clusters.

iii)   Measurements of cosmic microwave background anisotropies have determined that the spatial curvature parameter for the universe is [11]

$$\Omega_k = -0.0027 \pm 0.0039 \tag{4}$$

and subsequently $\Omega$, the universe's overall density parameter, is given by

$$\Omega = 1 - \Omega_k = 1.0027 \pm 0.0039 \tag{5}$$

Other theoretical considerations also seem to indicate that $\Omega$ must be almost exactly equal to 1.

iv)     Observations by Perlmutter et al [12] and Riess et al [13] of the light curves and redshifts of type Ia supernovae at high z indicate that the expansion of the universe is accelerating. The standard cosmological model density parameters for combined baryonic and dark matter, $\Omega_{M,O} = \Omega_{DM,O} + \Omega_{B,O}$, and dark energy, $\Omega_{\Lambda,O}$, that provide the best fit to the supernovae results, with the constraint that $\Omega = 1$, are [12];

$$\Omega_{M,O} = 0.28 \tag{6a}$$

$$\Omega_{\Lambda,O} = 0.72 \tag{6b}$$

and [13];

$$\Omega_{M,O} = 0.24 \tag{7a}$$

$$\Omega_{\Lambda,O} = 0.76. \tag{7b}$$

There are other observations that are consistent with the standard cosmological model, but the four listed above provide the foundation for the $\Lambda$CDM model.

\

The standard cosmological model does provide an explanation for the observations listed but it is far from satisfactory. In the opinion of this author, there are three major problems. First, no stable weakly interacting massive particles have been detected. If future experiments with the LHC come up empty it should be concluded that no such particles exist. Second, the cosmological constant or dark energy stems from observation iv listed above. However, if one steps back and asks what the observation initially implied, it is that the conventional dark matter theory is incorrect. Particles such as WIMPs cannot lead to an accelerating universe. The need to provide a cosmological constant or dark energy, in addition to dark matter, in order to explain the acceleration would seem to indicate a fundamental lack of understanding of what dark matter actually is. Third, the modeling of the expected dark matter distribution leads to a profile [14-15] that is at best only moderately successful in matching the observed rotational curves of spiral galaxies even with the distributions free parameters. Specifically, it does not lead to the rotational curves approaching a constant value in the outer regions. In addition, there is found an empirical relationship between $M_B$, the baryonic mass of a galaxy, and v, a galaxy's constant outer rotational velocity. This is referred to as the Baryonic Tully-Fisher Relationship (BTFR) [16-17]. The BTFR as given by McGaugh [17] is

$$M_B = A\ v^4 \tag{8a}$$

with

$$A = (47 \pm 6)\, M_\odot\, km^{-4}\, s^4. \tag{8b}$$

The BTFR is a relationship with surprisingly little scatter that ranges over five orders of magnitude of galactic baryonic masses. Conventional dark matter theory does not provide a natural explanation for the BTFR.

It is the third point, the inability of dark matter theory to explain the BTFR, which is the primary motivation behind alternatives to the dark matter hypothesis. For example, MOdified Newtonian Dynamics (MOND) as proposed by Milgrom [18-20] postulates that the inertia of an object varies with acceleration in a manner that specifically produces the BTFR. When applied, MOND does lead to better agreement with the galactic rotational curves. However, MOND doesn't explain much more than the BTFR and galactic rotation curves. In fact, the flatness of the universe does not naturally fall out of MOND and MOND still requires the existence of dark energy to explain observation iv, the acceleration of the expansion. As a result, there is a reluctance to accept MOND on par with the ΛCDM theory.

In a series of papers [1-4], the author presented the theory of an induced energy polarized vacuum as an alternative to both the current theory of dark matter and MOND. In this theory the gravitational field of a baryonic mass induces an energy contribution from the vacuum. The resulting distribution of the vacuum energy leads naturally to the BTFR [1], leads to excellent agreement with the rotational curves of galaxies [2,3] as well as the velocity dispersion and shear measurements taken with the Coma cluster [4]. The theory of an induced energy polarized vacuum is in agreement with the observations that are generally attributed to dark matter. The key points of this theory will be provided in Section 2

In Section 3 the author's theory will now be applied to superclusters. The average energy density of the induced vacuum contribution, from a model of the distribution of the superclusters, will be determined. It will be shown that the theory of an induced energy polarized vacuum is consistent with the observed flatness of the universe. The theory and supercluster model will then be used to determine the nature of the universe's expansion. It will be shown that the theory of an induced energy polarized vacuum is consistent with the acceleration of the expansion, without the need for dark energy. It must be stressed that there are no free parameters in the theory. In all cases, whether it is galactic rotation curves, velocity dispersions within clusters, or the vacuum contribution to the energy density of the universe, the input to the theory is the baryonic mass distribution.

## 2. Theory

Vacuum energy is conventionally taken to arise from the zero-point energy contributions associated with quantum fluctuations in the vacuum. This vacuum energy is typically taken to be the source of dark energy or the cosmological constant [21,22]. Unfortunately, calculations performed in quantum field theory on the expected vacuum energy density leads to an enormously large value that is over 120 orders of magnitude greater than that allowed by observations. The theory of an induced energy polarized of the vacuum as presented by the author does not treat vacuum energy in this conventional manner.

### 2.1 Hypothesis

Heisenberg's uncertainty principle allows for particle-antiparticle pairs to continually come into and out of existence in the vacuum. One of the particles has a positive energy while its antiparticle has a negative energy. The maximum lifetime of each of these particles, $\tau$, as given by the uncertainty principle is;

$$\tau \cong \frac{\hbar}{2\,|E|} \qquad\qquad (9)$$

where E is the energy of a given particle. When a particle interacts with an antiparticle in this sea of virtual particles both particles cease to exist. The total positive energy density and the total negative energy density due to these virtual particles will be enormous but the overall energy density of the vacuum, in the absence of an external gravitational field will be zero. However, in the presence of a gravitational field, the particles with positive energy will accelerate towards the gravitational source during their lifetime while the negative energy particles will accelerate away. In this case, it is found that the energy density of the vacuum surrounding a baryonic mass distribution will no longer be equal to zero.

### 2.2 Model

In order to proceed from the above hypothesis, the author considered the analogous situation of a dielectric in the presence of an electric field. Each of the individual virtual particles in the vacuum is treated as an energy dipole with its dipole moment given by

$$\mathbf{p_E} = E \langle \mathbf{x} \rangle_t \tag{10}$$

where $\langle \mathbf{x} \rangle_t$ is the time-averaged displacement of the particle towards or away from the gravitational source during its lifetime. The energy dipole moment density, $\mathbf{P_E}$, is then given by

$$\mathbf{P_E} = N \, \overline{\mathbf{p_E} \, t_E} \tag{11}$$

where N is the rate per unit volume at which the virtual particles (both positive and negative energy) come into existence, $t_E$ is the lifetime of a given particle ($t_E \leq \tau$), and the bar over $\mathbf{p_E}t_E$ represents an averaging over the particles. The resulting energy density of the vacuum, $\varepsilon_V$, surrounding a given gravitational field source is then given by

$$\varepsilon_V = -\nabla \cdot \mathbf{P_E} \, . \tag{12}$$

and the resulting induced gravitational field contribution, $\mathbf{g_V}$, due to this energy polarized vacuum will in turn be given by

$$\mathbf{g_V} = \frac{G}{c^2} \int_{V'} \frac{\varepsilon_V dV'}{(r-r')^3} \, (\mathbf{r} - \mathbf{r'}) \, . \tag{13}$$

In the case of spherical symmetry or in the far field limit where $r \rightarrow \infty$, (13) with the use of (12), simplifies to

$$\mathbf{g_V} = \frac{4\pi G}{c^2} \mathbf{P_E} \, . \tag{14}$$

To determine how the energy dipole moment density, $\mathbf{P_E}$, depends on the total gravitational field, the behavior of the virtual particles was treated semi-classically. The probability function for the distance travelled by a particle before interacting with an antiparticle was taken to be given by the standard Beer-Lambert law. From this model, the following relationship between the energy dipole moment density and the total gravitational field, g, was derived [1];

$$\mathbf{P_E} = \frac{c^2}{4\pi G} \, g_0 \left[ \frac{3}{2} \left( 1 - \frac{e^{-\gamma\left(\frac{g}{g_0}\right)}}{\gamma} \right) \right] \hat{\mathbf{g}} \tag{15}$$

where

$$\gamma = \frac{\sqrt{\pi}}{2} \frac{\text{erf}\left(\sqrt{\gamma\left(\frac{g}{g_0}\right)}\right)}{\sqrt{\gamma\left(\frac{g}{g_0}\right)}} \, . \tag{16}$$

The parameter $g_0$ in (15) and (16) can be expressed as

$$g_0 = \frac{8\pi G}{c^2} \frac{E}{\sigma} \tag{17}$$

with $\sigma$ being the cross-sectional area for a particle-antiparticle interaction. The expression for the energy dipole moment density, equations (15) and (16), can be expanded leading to

$$\mathbf{P_E} = \frac{c^2}{4\pi G} \, g_0 \left[\left(\frac{g}{g_0}\right) - \frac{3}{5}\left(\frac{g}{g_0}\right)^2 + \text{O}\left[\left(\frac{g}{g_0}\right)^3\right]\right] \hat{\mathbf{g}}. \tag{18}$$

Therefore to second order in the far field limit, $g \ll g_0$, it follows from (18) and (14) that

$$g_v = g - \left(\frac{3}{5g_0}\right) g^2. \tag{19}$$

Then by substituting $g = g_B + g_v$ into (19), where $g_B$ is the gravitational field due to the baryonic mass, and taking that in the far field $g_B \ll g_v$ it follows that

$$g_B = \left(\frac{3}{5g_0}\right) g^2. \tag{20}$$

which in turn leads to the BTFR,

$$M_B = \left(\frac{3}{5Gg_0}\right) v^4. \tag{21}$$

The BTFR is therefore a natural consequence of the theory. By equating the coefficients of (8a) and (21),

$$g_0 = \frac{3}{5AG} \, , \tag{22}$$

and for the value of A as given in (8b) the parameter $g_0$ is then given by

$$g_0 = (9.6 \pm 1.2) \text{ x } 10^{-11} \text{ m s}^{-2}. \tag{23}$$

From the determined value for $g_0$ some of the details of the virtual particles involved can be found for the given model [1]. For the value for the BTFR coefficient as given by (8b), the following inequalities are obtained;

$$\sigma < 2.7 \text{ x } 10^{-45} \text{ m}^2, \tag{24a}$$

$$|E| < 4.2 \text{ x } 10^{-29} \text{ J } (2.6 \text{ x } 10^{-10} \text{ eV}). \tag{24b}$$

The only known interaction between particles that has a cross-section as small as given by (24a) is the weak interaction between leptons [27]. The only known lepton that could possibly have such a small energy as given by (24b) is the electron neutrino. The theory of an induced energy polarized vacuum therefore indicates that the dominant contributors to the induced gravitational field surrounding a baryonic mass are virtual electron neutrinos and their antiparticles. Of course, it is also possible that a currently unknown particle is responsible.

## 3. Comparisons with observations attributed to dark matter

From equations (15) and (12) the distribution of the induced vacuum energy can be determined. Then, by (13), the resulting contribution that this vacuum energy makes to the total gravitational field surrounding any baryonic mass distribution can be determined. The theory of an induced energy polarized vacuum was first applied to the Galaxy [2]. Using observation based models of the baryonic mass distributions of the bulge and disc of the Galaxy, a theoretical rotational velocity curve was determined. There was found to be very good agreement between the theoretical and the observed rotational curve. Then the theory was applied to the general rotational curves of modeled spiral galaxies [3]. It was found that the theory readily produced the features seen in real rotation curves. The theory was then applied to the Coma cluster [4]. From a model of the baryonic mass distribution of the cluster, the theoretical virial mass was determined and was found to be in good agreement with previous virial mass estimates. In addition, the theoretical velocity dispersions for the galaxies of the cluster were in good agreement with observations. Theoretical shear values for the Coma cluster were also found to be overall in good agreement with observations. Therefore the theory of an induced energy polarized vacuum is in excellent agreement with the observations that are currently attributed to dark matter.

Again, it needs to be stressed that no free parameters are involved for these applications of the theory. The only parameter in the model, $g_0$, is determined by the coefficient of the BTFR. This coefficient and the baryonic mass distribution are the only inputs required in the determination of the dynamics of galaxies and clusters. This is in contrast to dark matter profiles which involve free parameters that are fitted to each individual galaxy or cluster.

## 3. Resulting Cosmology

## 3.1 Density parameter for the vacuum contribution

In order to determine the contribution of the induced energy polarized vacuum to the overall energy density of the universe, a large-scale model of the baryonic mass distribution of the universe is needed. Observations show that the galaxies in the universe are not uniformly distributed but are typically found in groups and clusters. These groups and clusters are in turn part of even larger structures called superclusters that surround large sparsely populated regions called voids. From a catalog of superclusters out to z ≤ 0.08 [24], it is estimated that the mean separation between superclusters is $D_{SC} \cong 100h^{-1}$ Mpc [25]. Observations of the redshift distribution of galaxies in narrow pencil-beam surveys [26] in turn indicate an apparent periodic distribution of galaxies with a regular separation of $128h^{-1}$ Mpc. This distribution was later shown to originate from the intersection of the narrow-beams with the tails of large superclusters [27]. Distances between high-density regions across the voids were also determined using narrow pencil beam surveys [28]. The median distances for the different samples ranges from $116h^{-1}$ Mpc to $143h^{-1}$ Mpc. From these listed results the current average separation of superclusters will be estimated to be

$$D_{SC,O} = (120 \pm 20)h^{-1} \text{ Mpc.} \tag{25}$$

Also this separation is approximately equal to the diameter of an average void [28-30].

As a simplified model of the baryonic mass distribution of the universe, it will be taken that the baryonic mass is lumped together at the location of uniform sized superclusters currently separated by $D_{SC,O}$. Each model supercluster will be taken to be the dominant source of the gravitational field for distances within $R_{SC,O}$ of the superclusters centre where

$$R_{SC,O} = D_{SC,O}/2 = (60 \pm 10)h^{-1} \text{ Mpc.} \tag{26}$$

The total baryonic mass associated with these modeled superclusters will be such that the average baryonic density within $R_{SC,O}$ is as given by (3), i.e. the model is set to agree with observation i. This leads to each modeled supercluster having a total baryonic mass of approximately $1 \times 10^{26} M_{sun}$. Although this supercluster model is very simple, in terms of the theory the dominant contribution that the induced energy polarized vacuum will make to the overall energy of the universe will come from the voids. It is the behavior and values of the gravitational fields within the voids which is of greatest importance. In this case, the above model of baryonic mass, localized at supercluster positions, would result in gravitational fields within the voids that would be expected to be in reasonable agreement with actual values.

With the given model it is a straightforward process to determine the total equivalent mass, $M_V$, that the induced vacuum contributes to a region of radius $R_{SC}$ surrounding a model supercluster. Substituting $v^2 = GM_V R_{SC}^{-1}$ into the BTFR as given by (21) results in

$$M_V = \left(\frac{5g_0}{3G}\right)^{1/2} M_B^{1/2} R_{SC} \tag{27}$$

Equation (27) can also be expressed in terms of the average baryonic and average vacuum energy densities, i.e.

$$\varepsilon_v = \left(\frac{5g_0 c^2}{4\pi G}\right)^{1/2} \left(\frac{\varepsilon_B}{R_{sc}}\right)^{1/2}. \tag{28}$$

Expressing (28) in terms of the density parameters for the baryonic mass, $\Omega_B = \varepsilon_B/\varepsilon_c$, and the induced vacuum contribution, $\Omega_V = \varepsilon_V/\varepsilon_c$, then results in

$$\Omega_V = \frac{1}{H}\left(\frac{10g_0}{3}\right)^{1/2}\left(\frac{\Omega_B}{R_{sc}}\right)^{1/2}. \tag{29}$$

Substituting from (1) and (26), the current value of the density parameter for the induced vacuum contribution will therefore be given by

$$\Omega_{V,O} = (0.57 \pm 0.11)h^{-3/2}. \tag{30}$$

For the value of $h = 0.738 \pm 0.024$, as determined by (2), the current density parameter for the induced vacuum contribution is

$$\Omega_{V,O} = 0.90 \pm 0.22 \tag{31}$$

and therefore by (31) and (3) the overall density parameter of universe as given by the theory of an induced energy polarized vacuum is

$$\Omega = \Omega_{V,O} + \Omega_{B,O} = 0.94 \pm 0.23. \tag{32}$$

The baryonic mass of the universe plus the resulting induced energy contribution from the vacuum is consistent with $\Omega$ being equal to 1. No additional contributor to the energy content of the universe, such as dark energy, is required to explain observation iii. At this time I can offer no explanation of why the induced energy contribution of the vacuum surrounding superclusters leads to $\Omega = 1$. Once again, it must be stressed that there are no free parameters in the theory, it is the distribution and mass of the superclusters that leads to this value for $\Omega$.

## 3.2 Expansion rate

Of course the primary reason for invoking the cosmological constant or dark energy was to explain the acceleration of the expansion of the universe. In the standard $\Lambda$CDM model both baryonic matter and dark matter are treated as nonrelativistic matter, $\Omega_{M,O} = \Omega_{DM,O} + \Omega_{B,O}$. Neglecting the radiation contribution, which is only of importance for the scale factor $a < 10^{-3}$, the Freidmann equation for the standard cosmological model becomes

$$\dot{a}^2 = H_o{}^2 \left( \frac{\Omega_{M,O}}{a} + \Omega_{DE,O} a^2 \right) \tag{33}$$

For the estimated values of $\Omega_{M,O}$ and $\Omega_{DE,O}$ as given either by Perlmutter et al [12], equations (6a-b), or by Reiss et al [13], equations (7a-b), the dark energy term has been dominating since $a > 0.7$ and therefore for the standard cosmological model the expansion of the universe is currently accelerating.

In the theory of an induced energy polarized vacuum there are two constituents (again neglecting the radiation contribution) that affect the expansion rate, the baryonic mass and the contribution induced from the vacuum. Substituting

$$R_{SC} = a \, R_{SC,O} \tag{34}$$

and

$$\varepsilon_B = \varepsilon_{B,O} \, a^{-3} \tag{35}$$

into (28) results in the following dependence of the average energy density of the vacuum contribution on the scale factor;

$$\varepsilon_v = \left( \frac{5 g_0 c^2 \varepsilon_{B,O}}{4 \pi G R_{sc,o}} \right)^{1/2} a^{-2}. \tag{36}$$

This dependence that $\varepsilon_v$ has on the scale factor leads to a negative pressure and the resulting Freidmann equation for the theory of an induced energy polarized vacuum theory is given by

$$\dot{a}^2 = H_o{}^2 \left( \frac{\Omega_{B,O}}{a} + \Omega_{V,O} \right). \tag{37}$$

For the purposes of this section, the density parameters given by (3) and (31) will be normalized so that $\Omega_{B,O} + \Omega_{V,O} = 1$, i.e. $\Omega_{B,O} = 0.039$ and $\Omega_{V,O} = 0.961$. For these values (37) shows that the induced vacuum contribution has been dominating since $a > 0.04$ and with $\dot{a}$ asymptotically approaching a constant value of $H_O \Omega_{V,O}{}^{1/2}$. So, as determined by the theory of an induced energy polarized vacuum, at the present time the universe is coasting. However, the model of the baryonic mass distribution of the universe presented assumes that the superclusters are bound systems, i.e. the size of the superclusters is not changing. If superclusters are in general unbound systems then the size and separation of the superclusters would be expected to evolve over time. To demonstrate the possible effect of an evolving supercluster distribution, consider the simple case where $R_{SC}$ stays constant in time. This will correspond to the average diameter of the voids remaining constant. A constant $R_{SC}$ will also equate to the number of modeled superclusters increasing in time as the baryonic mass of each supercluster decreases, i.e. the universal expansion is pulling the superclusters apart. The average baryonic energy density would continue to follow (35) with the current density parameter as given by (3).

For the case of constant $R_{SC}$, by (28) and (35)

$$\varepsilon_v = \left(\frac{5g_0c^2\varepsilon_{B,o}}{4\pi GR_{sc,o}}\right)^{1/2} a^{-3/2} \tag{38}$$

and therefore the Freidmann equation for this unbound supercluster model then becomes

$$\dot{a}^2 = H_o{}^2 \left(\frac{\Omega_{B,o}}{a} + \Omega_{V,O}a^{1/2}\right). \tag{39}$$

For an increasing scale factor, equation (39) shows that the induced vacuum contribution term now leads to a positive acceleration of the expansion of the universe. For values of $\Omega_{B,O} = 0.039$ and $\Omega_{V,O} = 0.961$ the induced vacuum contribution in this model has been dominating since a > 0.12. Therefore, for this unbound supercluster model, the induced energy polarized vacuum leads to an accelerating expansion of the universe without the need for dark energy.

The observational evidence for an accelerating universe comes from the distance modulus versus redshift values for type Ia supernovae. Figure 1 shows the fitted theoretical relationships between the distance modulus and the redshift for the standard $\Lambda$CDM cosmological model for both the density parameters of Perlmutter et al [12], equations (6a-b), and Reiss et al [13], equations (7a-b). Also, shown on the figure are the theoretical relationships between distance moduli and redshift for both the bound and unbound supercluster models of the induced energy polarized vacuum theory. All four of the models are plotted with respect to the case of a coasting universe, i.e. $\dot{a} = 0$. Included on Figure 1 are the actual type Ia supernovae results [12-13]. As is shown in the figure, the unbound supercluster model leads to an accelerating expansion that is very compatible with the type Ia supernovae results. Therefore an induced energy polarized vacuum therefore can explain the observations not only attributed to both dark matter but also to those attributed to dark energy.

## 4. Conclusion

The theory of an induced energy polarized vacuum provides an alternative to the standard cosmological model. The theory has previously been shown to lead to the BTFR [1], to agree with the observed rotation curve of the Galaxy [2], to readily produce the observed features in the rotational curves of other spiral galaxies [3], and to agree with observations of the Coma cluster [4]. All without the need for any free parameters. The theory of an induced energy polarized vacuum is now applied to superclusters. From a model of the distribution of superclusters, the overall density parameter of universe as given by the theory of an induced energy polarized vacuum is determined to be $\Omega = 0.94 \pm 0.23$. This is consistent with a geometrically flat universe. In addition, by treating the superclusters as unbound systems, the theory leads to an accelerating expansion of the universe in line with observations and without requiring dark energy.

The theory of an induced energy polarized vacuum is a variation of the standard dark matter model in the sense that non-baryonic particles, i.e. virtual electron neutrino particle-antiparticle pairs, provide a real energy density that contributes to the gravitational field surrounding a given baryonic mass distribution. No changes in gravitational theory or Newtonian mechanics are required. However, unlike the standard theory of dark matter, the additional contribution provided is directly coupled to the baryonic mass. Knowing the baryonic mass distribution is all that is required to determine the rotational curve for a galaxy or the dynamics of a cluster. In addition, unlike the standard theory of dark matter, the BTFR is a natural consequence of the theory of an induced energy polarized vacuum. The theory of an induced energy polarized vacuum may also be seen as a variation of the standard dark energy model in the sense that in the theory the vacuum provides an energy contribution with an associated negative pressure. However, unlike the theory of dark energy, the vacuum contribution is directly coupled to the baryonic mass. Knowing the baryonic mass distribution of the universe will determine its geometry and the nature of its expansion.

What is especially appealing about the theory of an induced energy polarized vacuum is the role of the baryonic mass. The induced vacuum energy contributes to the gravitational field surrounding a baryonic mass and to to the energy content of the universe, but this contribution is solely dependent on the baryonic mass distribution; a universe without baryonic mass would truly be empty. Astronomical observations that lead to the determination of the baryonic mass distribution will lead directly to the determination of the dynamics of galaxies, clusters, superclusters, and the expansion of the universe as a whole.

## 4. Discussion

Although readers may abject the model of particle-antiparticle interactions used in the theory, the specific model will have little effect on the results. If quantum fluctuations in the vacuum are affected in any manner by an external gravitational field then an equivalent energy dipole moment density can be defined. In general such a $\mathbf{P_E}$ can be expanded as per (18) which will lead to the BTFR, and with the observed BTFR coefficient this will lead to a vacuum energy distribution that will result in the same galactic rotational curves, cluster dynamics, energy density, and nature of expansion that the current model provides. Different models will affect the relationship between $g_0$ and the parameters of the particular model. In the model provided this leads to the relation as given by (17) which in turn leads to the values for the particles as given by (24a-b).

Also the different models will impact the strong field limit, $g \gg g_o$. For the model presented, and in the case of spherical symmetry, by (14) and (15) the vacuum contribution to the gravitational field is found to saturate at a value of

$$g_v \rightarrow \frac{3}{2} g_o = (1.4 \pm 0.2) \text{ x } 10^{-10} \text{ m s}^{-2}. \tag{40}$$

The Sun's gravitational field greatly exceeds $g_o$ throughout the solar system. As such, if the author's theory and model is extended to the solar system, the value of $g_v$ would be found to be at the saturated value of $(1.4 \pm 0.2)$ x $10^{-10}$ m s$^{-2}$. Although the recent Pioneer anomaly results of Turyshev et al [31] found that the bulk of the anomalous acceleration can be explained by the emission of thermal radiation by the Pioneer vehicle, by using their Figure 3 the anomalous acceleration for a distance $\geq 20$ AU is estimated at

$$a_{\text{PIONEER}} = (1.4 \pm 1.9) \times 10^{-10} \text{ m s}^{-2}, \tag{41}$$

with the given uncertainty being solely due to the modeling of the thermal term. The value given in (40), derived from the theory of an induced energy polarized vacuum theory, is actually in very good agreement with these latest results, although the absence of any observed anomaly on the order of (40) within the inner solar system ($\leq 15$ AU) still needs to be addressed. Be that as it may, the theory of an induced energy polarized vacuum is testable in that it predicts that at some distance from the Sun an anomalous acceleration on the order of (40) must exist.

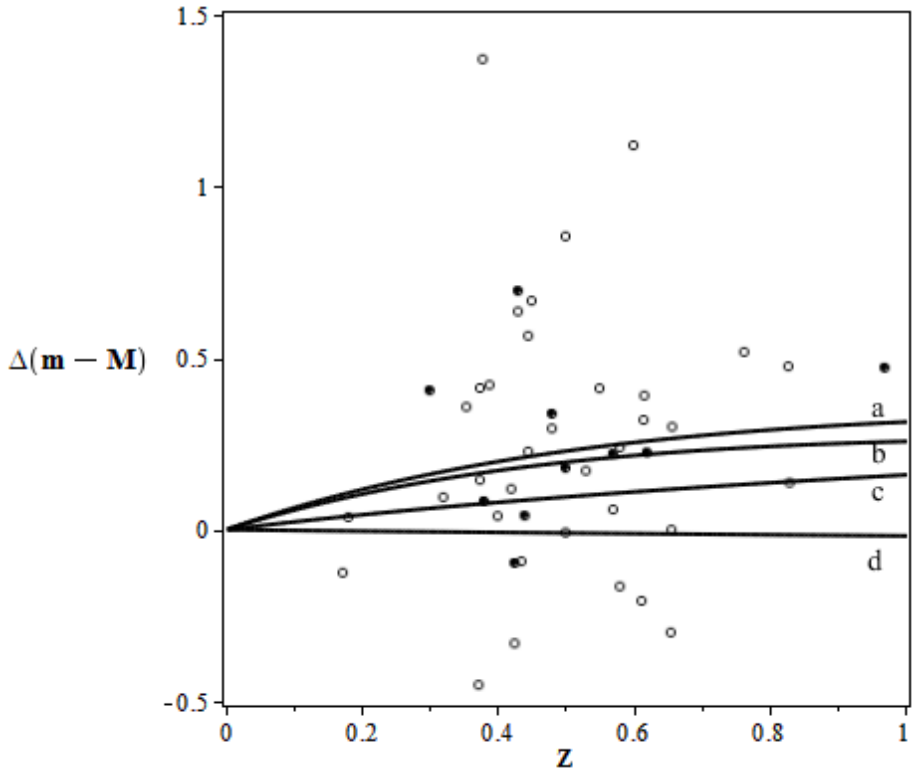

**Figure 1:** Distance modulus versus redshift (z) for the different models with Δ(m−M) being the difference between the values and the prediction for a coasting universe: a) standard cosmological model using the values given in (6a-b) [12], b) standard cosmological model using the values given in (7a-b) [13], c) induced energy polarized vacuum values for unbound superclusters, d) induced energy polarized vacuum values for bound superclusters. Also included on the figure is the type Ia supernovae results from (o) - Perlmutter et al [12] and (●) - Riess et al [13]